\documentclass[useAMS,usenatbib]{mn2e}
\usepackage{graphicx}
\usepackage{url}
\def\apj{ApJ}%
\def\apjl{ApJ}%
\def\apjs{ApJS}%
\def\aap{A\&A}%
\def\mnras{MNRAS}%
\title[The lower limits of disc fragmentation]
{The lower limits of disc fragmentation and the prospects for observing fragmenting discs}
\author[Stamatellos et al. ]
{Dimitris Stamatellos$^{\!\  1,}$\thanks{E-mail:D.Stamatellos@astro.cf.ac.uk}, Ana\"elle~Maury$^{\!\  2,3}$, Anthony~Whitworth$^{\!\  1}$, Philippe   Andr\'e$^{\!\  3}$ \\
 $^1$ School of Physics \& Astronomy,Cardiff University, Cardiff, CF24 3AA, Wales, UK\\
 $^2$ European Southern Observatory, Karl Schwarzschild Str. 2, 85748, Garching bei M\"unchen, Germany\\
$^3$ Laboratoire AIM, CEA/DSM-CNRS-Universit\'e Paris Diderot, IRFU/Service d' Astrophysique, C.E. Saclay, Orme des \\Merisiers, 91191 Gif-sur-Yvette, France\\}

\begin{document}

\date{Accepted 2010 . Received 2010 October 14; in original form 2010 October 14}

\pagerange{\pageref{firstpage}--\pageref{lastpage}} \pubyear{201-}

\maketitle

\label{firstpage}

\begin{abstract}
A large fraction of brown dwarfs and low-mass hydrogen-burning stars may form by gravitational fragmentation of protostellar discs.  We explore the conditions for disc fragmentation and we find that they are satisfied when a disc is large enough ($\stackrel{>}{_\sim}100$~AU) so that its outer regions can cool efficiently,  {\it and} it has enough mass to be gravitationally unstable, at such radii. We perform radiative hydrodynamic simulations and show that even a disc with mass $0.25~{\rm M_{\sun}}$ and size 100~AU  fragments. The disc mass, radius, and the ratio of disc-to-star mass ($M_{\rm D}/M_{\star}\approx0.36$) are smaller than in previous studies (Stamatellos \& Whitworth 2009a). We find that fragmenting discs drastically decrease in mass and size within a few $10^4$ yr of  their formation, since a fraction of their mass, especially  outside $\sim\!100$~AU is consumed by the new stars and brown dwarfs that form. Fragmenting discs  end up  with masses  $\sim\!0.001-0.1\ {\rm M}_{\sun}$, and  sizes  $\sim\!20-100$~AU. On the other hand, discs that are marginally stable evolve on a viscous timescale, thus living longer ($\sim\!1-10$ Myr). We produce simulated images of fragmenting discs and find that observing discs that are undergoing fragmentation is possible using current  (e.g. IRAM-PdBI) and future (e.g. ALMA) interferometers, but highly improbable due to the short duration of this process. Comparison with observations shows that many observed discs may be remnants of discs that have  fragmented at an earlier stage. However, there are only a few candidates that are possibly massive and large enough to currently be gravitationally unstable.  The rarity  of massive ($\stackrel{>}{_\sim}\!0.2\  {\rm M}_{\sun}$), extended ($\stackrel{>}{_\sim}100$ ~AU) discs indicates either  that such discs are highly transient (i.e. form, increase in mass becoming gravitationally unstable due to infall of material from the surrounding envelope, and quickly fragment), or  that their formation is suppressed (e.g. by magnetic fields). We conclude that current observations of early-stage discs cannot exclude the mechanism of disc fragmentation.

\end{abstract}


\section{Introduction}

The formation of low-mass hydrogen-burning stars, brown dwarfs (BDs) and planetary-mass objects is a critical problem, since it addresses  the question of why there is a minimum mass for star formation.  These objects are difficult to form by gravitational fragmentation of unstable gas, as for masses in the BD regime ($\stackrel{<}{_\sim}80~{\rm M}_{\rm J}$, where ${\rm M}_{\rm J}$ is the mass of Jupiter) a high density is required for the gas to be Jeans unstable. Thus, if BDs form according to the standard model of star formation, i.e. the collapse of a prestellar core,  then the pre-BD core must have a density of $\stackrel{>}{_\sim}10^{-16} {\rm g\ cm}^{-3}$. 
 Padoan \& Nordlund (2004)  and Hennebelle \&  Chabrier (2008) suggest that these high density cores may form by colliding flows in a turbulent magnetic medium. However, it has not been demonstrated that this mechanism works when the thermodynamics and hydrodynamics are properly simulated, and there is no obvious reason why it should reproduce the observed binary properties of BDs and low-mass stars. 
 
An alternative way to reach the high densities required for the formation of BDs is in protostellar discs (Whitworth \& Stamatellos 2006; Stamatellos et al. 2007b). These discs form around newly born stars and grow quickly in mass by accreting material from the infalling envelope (Terebey et al. 1984; J{\o}rgensen et al. 2009; Vorobyov \& Basu 2010; Machida \& Matsumoto 2010; Walch et al. 2010). They become unstable and fragment if the matter infalling on them cannot efficiently redistribute its angular momentum outwards in order to accrete onto the central star (Attwood et al. 2009). Whitworth \& Stamatellos (2006) argue that fragmentation happens in the outer disc where the cooling time is short enough (Matzner \& Levin 2006; Boley 2009). Stamatellos et al. (2007b) show that this is a viable mechanism, and Stamatellos \& Whitworth (2009a)  determine the statistics of the low-mass objects produced by this mechanism, performing an ensemble of radiative hydrodynamic simulations of fragmenting discs. The disc fragmentation model reproduces the critical constraints set by the observed statistical properties of low-mass objects, and these have not yet been explained by other formation mechanisms. More specifically the model reproduces the shape of the mass distribution of low-mass objects (IMF), the lack of BDs  as close companions to Sun-like stars (i.e. the BD desert), the presence of discs around BDs, the statistics of low-mass binary systems, and the formation of free-floating planetary mass objects (Stamatellos \& Whitworth 2009a). 

The main argument against the formation of low-mass stars and BDs by fragmentation of discs is whether the massive (a few times 0.1 M$_{\sun}$), extended discs ($\stackrel{>}{_\sim}100$~AU) assumed by this mechanism can be realized in nature. From a theoretical point of view the formation of such discs is inevitable considering the angular momentum content of their parental cores. For example, a $1.4\,{\rm M}_\odot$ prestellar core with ratio of rotational to gravitational energy $\beta\equiv{\cal R}/|\Omega|$ will -- if it collapses monolithically -- form a protostellar disc with outer radius $R_{_{\rm DISC}}\sim 400\,{\rm AU}\,(\beta/0.01)$. The observations of Goodman et al. (1993) indicate that many prestellar cores have $\beta \sim 0.02$, hence the formation of extended, self-gravitating discs should be rather common.

 Recent numerical studies reach mixed conclusions on this topic. Simulations that ignore the effects of magnetic fields (Attwood et al. 2009; Machida \& Matsumoto 2010) find that such large discs form frequently in simulations of turbulent  and/or rotating molecular clouds. The presence of magnetic fields is expected to  act against the formation of a centrifugally supported disc because angular momentum is removed by magnetic effects (e.g. magnetic braking, outflows). However, it is uncertain whether, in realistic conditions, magnetic fields can totally suppress the formation of self-gravitating discs. Hennebelle \& Fromang (2008) and Hennebelle \& Teyssier (2008) find that in the ideal MHD approximation the formation of a disc is suppressed if the magnetic field is strong enough (with strength similar to what is inferred by observations, e.g. Crutcher et al. 2004, Falgarone et al. 2008) and parallel to the rotation axis of the collapsing star-forming core. If the magnetic field is mis-aligned then its effect is weakened and discs may form, but  fragmentation seems to be suppressed (Hennebelle \& Ciardi 2009). These results are supported by ideal MHD simulations that include the effects of radiative transfer (Commer{\c c}on  et al. 2010). Disc formation and fragmentation in strongly magnetized clouds appears to be possible only for rapidly rotating clouds (Machida et al. 2005). Ambipolar diffusion tends to weaken the effects of the magnetic fields, but for realistic magnetic field strengths and ionization rates not enough to limit the effect of magnetic breaking (Mellon \& Li 2008, 2009). The situation changes in resistive MHD calculations. Ohmic dissipation is probably important at high densities and may dominate over other processes at densities $10^{12}$~cm$^{-3}$ (Nakano et al. 2002). Machida et al. (2008, 2010) using a resistive MHD method, for a wide range of rotational-to-magnetic energies, find that disc formation is possible. In these simulations fragmentation happens either on a large scale and produces wide binaries, or after the formation of the first core and produces close binaries. Krasnopolsky et al. (2010) find that a resistivity on the order of $10^{19}$~cm$^2$~s$^{-1}$ is needed for the formation of discs with sizes on the order of  100 AU.

The goal of this paper is to determine the lower limits of fragmentation in terms of disc mass and size, and compare these with observations of early-stage discs. Two conditions have to be satisfied for a disc to fragment: (i)~the disc must be Toomre unstable i.e. it has to be massive enough so that gravity can overcome thermal and local centrifugal support (Toomre 1964), and (ii)~the disc must  cool efficiently so that proto-condensations forming in the disc do not simply undergo an adiabatic bounce and dissolve (Gammie 2001). These criteria can be satisfied for discs considerably smaller in mass and size than the discs assumed by Stamatellos \& Whitworth (2009a). In particular,  the fragmentation criteria are  fulfilled  for discs that  (i) are large enough so that their outer regions can cool fast enough, i.e. for discs with radii $\stackrel{>}{_\sim} 100$~AU,  and (ii) have enough mass (i.e. an appropriate surface density profile) so that they are Toomre unstable at such radii. In this paper we present an ensemble of simulations exploring the possibility of fragmentation of small, marginally unstable discs. We find  that even discs with masses 0.25 M$_{\sun}$ and sizes 100 AU fragment. We produce simulated images of such discs that show that they are observable by current facilities (e.g the IRAM Plateau de Bure interferometer; hereafter PdBI), although they should  be rarely observed as they quickly fragment. Comparison with observations shows that a few marginally unstable discs may exist. Additionally, many discs are consistent with being remnants of disc that have fragmented in the past. 

The paper is structured as follows. In Section 2 we describe the initial conditions of the disc simulations, the computational method used, and the results regarding the lower limits of disc fragmentation as they are derived by the simulations. In Section 3 we produce synthetic images of discs during the fragmentation process, and  in Section 4 we  compare the properties of fragmenting discs with observations of candidate early-stage discs. Finally, in Section 5 we summarise this work and  discuss its implications.

\section{Hydrodynamic simulations of marginally unstable discs}
\subsection{Disc initial conditions: Density and temperature profiles}
\label{subsec:ics}
We assume a star-disc system in which the central primary star  has initial mass $M_\star=0.7\,{\rm M}_{\sun}$.  The initial disc mass, 
$M_{_{\rm D}}^i$, and the initial disc radius, $R_{_{\rm D}}^i$, are free parameters. The disc mass  is varied from 0.15 to 0.30 ${\rm M}_{\sun}$ and the disc radius from 60 to 150~AU (see Table~\ref{tab:runs}). The initial surface density is 
\begin{equation}
\Sigma_{_0}(R)=\Sigma(1 {\rm AU})\,\left(\frac{R}{\rm AU}\right)^{-p}\,,
\end{equation}
and the disc temperature
\begin{equation}\label{EQN:TBG}
T_{_0}(R)=250\,{\rm K}\,\left(\frac{R}{\rm AU}\right)^{-q}+10~{\rm  K}\,,
\end{equation}
where $\Sigma(1 {\rm AU})$ is determined by the disc mass and radius, $p$ and $q$ are free parameters, and $R$ the distance from the central star measured on the disc midplane. The values of $q$ and $p$ are rather uncertain and may evolve with time. Andrews et al. (2009) observed a sample of circumstellar discs in Ophiuchus and estimated that the disc surface density drops with an exponent $p\approx 0.4-1.0 $. In Taurus-Auriga and Ophiuchus-Scorpius star formation regions  Andrews \& Williams (2007) find a  median $p\approx 0.5$; they also find a temperature profile with  $q\approx 0.4-0.74$. Osterloh \& Beckwith (1995) find that the disc temperature drops with an exponent $q\approx 0.35-0.8$. 

Previous studies (e.g. Stamatellos \& Whitworth 2009a) have used $p=1.75$ and $q=0.5$, resulting in a disc with almost uniform Toomre parameter $Q$. The conditions  are more favourable for fragmentation for lower values of $p$ (as the disc has more mass at larger radii), and higher values of $q$ (as the temperature drops faster with the distance from the central star). In this study we set $p=1$, hence there is equal mass per unit radius in the disc, and $q=0.75$ or $q=0.5$ (see Table~\ref{tab:runs}). These parameters make fragmentation easier than in previous simulations (Stamatellos et al. 2007; Stamatellos \& Whitworth 2009a), but  they are not extreme values when compared with disc observations.

\begin{table}
\begin{minipage}{\columnwidth}
\caption{Parameters of the simulations of marginally unstable discs.  M$_{\rm D}^i$: disc initial mass,  $R_{\rm D}^i$: disc initial radius, M$_{\rm D}^f$: disc final mass (i.e. at $t=40$~kyr),  $R_{\rm D}^f$: disc final radius, $p$: exponent of the disc surface density profile,  $q$: exponent of the disc temperature profile,  $N_{\rm f}$: number  of objects formed. The colour  corresponds to Figs.~\ref{fig:nonfragment}, \ref{fig:fragment}.}
\label{tab:runs}
\centering
\renewcommand{\footnoterule}{}  
\begin{tabular}{@{}clcccccclc} \hline
\noalign{\smallskip}
id 	&	M$_{\rm D}^i$  & $R_{\rm D}^i$ & M$_{\rm D}^f$  & $R_{\rm D}^f$ & $p$ & $q$ & $N_{\rm f}$ &colour \\
 	&	(M$_ {\sun}$) & (AU) & (M$_ {\sun}$) &  (AU) &  &  &  & \\
\noalign{\smallskip}
\hline
\noalign{\smallskip}
1  		&	0.3	& 150 & 0.002	&20		&  1	& 0.75	 & 3 & black\\ 
2  	 	&	0.3 	& 100 &  0.25    &250	& 1 	& 0.5 	 & - & black\\
3   		&	0.3  	& 100 &  0.03	&35		&  1 	& 0.75	& 4  & green \\ 
4   	 	&	0.2  	& 100 & 0.17	&240	&  1	& 0.75	 & -  & purple\\
5  		&	0.25	& 100 & 0.07	&70		&  1	& 0.75	 &  1  &red\\
6 		&	0.25	&   80 & 0.17	&240	&  1	& 0.75	 & -  &green\\ 
7 		&      0.25	&   60 & 0.17	&220	&  1	& 0.75	 & -  &cyan\\ 
8 		&	0.2	&   80 & 0.16	&220	&  1	& 0.75	 & -  & red\\ 
9  		&	0.15	&   90 & 0.12	&220	&  1	& 0.75	 & -  & blue\\ 
 \noalign{\smallskip}
\hline
\end{tabular}
\end{minipage}
\begin{minipage}{\columnwidth}
\caption{The properties of the objects formed by disc fragmentation. We list their final mass ($M_{\rm f}$), formation radius ($r_{\rm in}$), final status (i.e. whether they remain bound to the central star or are ejected from the disc), and their final semi-major axis ($a_{\rm fin}$) and eccentricity ($e$) (if they are still bound to the central star), or their velocities ($v_{\rm ej}$)  (if they are ejected from the disc). }
\label{tab:statistics}
\centering
\renewcommand{\footnoterule}{}  
\begin{tabular}{@{}ccccccc} \hline
\noalign{\smallskip}
id 	&$M_{\rm f}$  & $r_{\rm in}$  & status & $a_{\rm fin}$  &  $e$ & $v_{\rm ej}$ \\
 	& (M$_ {\sun}$) & (AU) & &  (AU) &   & (km s$^{-1}$) \\
\noalign{\smallskip}
\hline
\noalign{\smallskip}
1	& { 0.10}   & 133 & bound & 37& 0.35 & -  \\
     	& { 0.060} &110 & bound &310& 0.22 &-\\
    	& { 0.059} &180 & bound &1250 & 0.97&- \\ 
\hline
3   	& { 0.116}&100 & bound &83& 0.47&-\\
  	&{ 0.023} &175& ejected &- &- & 2.4 \\
	&{0.018}  &200& bound &170 & 0.41&-\\
	& { 0.006}&140& ejected &- &- & 1.9 \\ 
	\hline
5  	& { 0.092}&70& bound&130& 0.24& -\\ 
 \noalign{\smallskip}
\hline
\end{tabular}
\end{minipage}
\end{table}

\subsection{Computational method}

We use the SPH code {\sc seren} (Hubber et al. 2011), which invokes an octal tree (to compute gravity and find neighbours), adaptive smoothing lengths, multiple particle timesteps, and a second-order Leapfrog integration  scheme. The code uses time-dependent artificial viscosity (Morris \& Monaghan 1996) with parameters $\alpha^\star=0.1$, $\beta=2\alpha$ and a Balsara switch (Balsara 1995), so as to reduce artificial shear viscosity. 

The radiative processes that regulate the disc thermodynamics  affecting the fragmentation process are treated with the Stamatellos et al. (2007a) diffusion method that includes radiation heating from the central protostar as described in Stamatellos et al. (2009b) and in Forgan et al. (2009). The method takes into account compressional heating, viscous heating, radiative heating by the background, and radiative cooling. It performs well, in both the optically thin and optically thick regimes, and has been extensively tested (Stamatellos et al. 2007a; Forgan et al. 2009). In particular it reproduces the detailed 3D results of Masunaga \& Inutsuka (2000), Boss \& Bodenheimer (1979), Boss \& Myhill (1992), Whitehouse \& Bate (2006),  and also the analytic results of Spiegel (1957) and Hubeny (1990).

\begin{figure*}
\centerline{
\includegraphics[width=0.9\columnwidth]{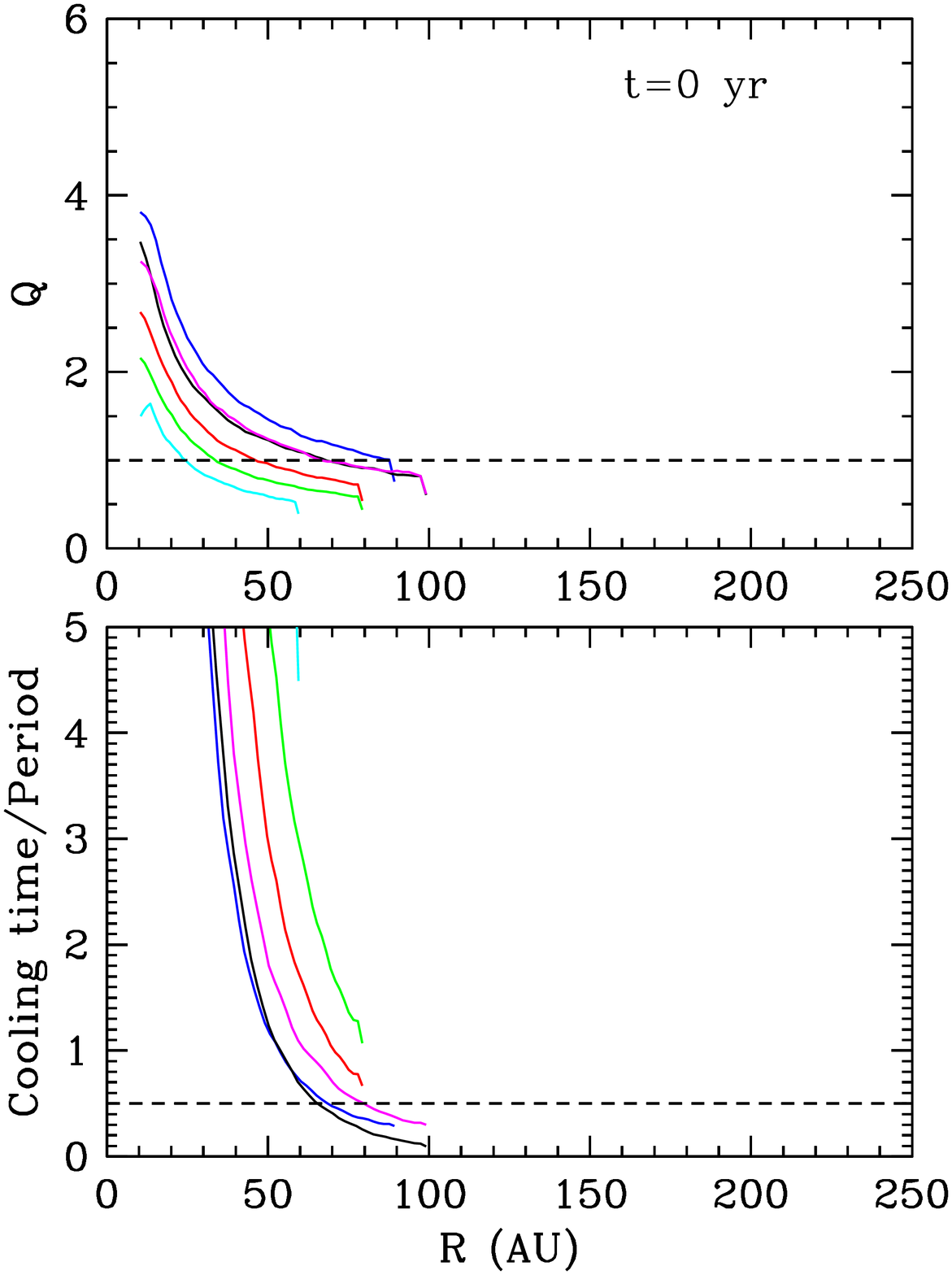}
\includegraphics[width=0.9\columnwidth]{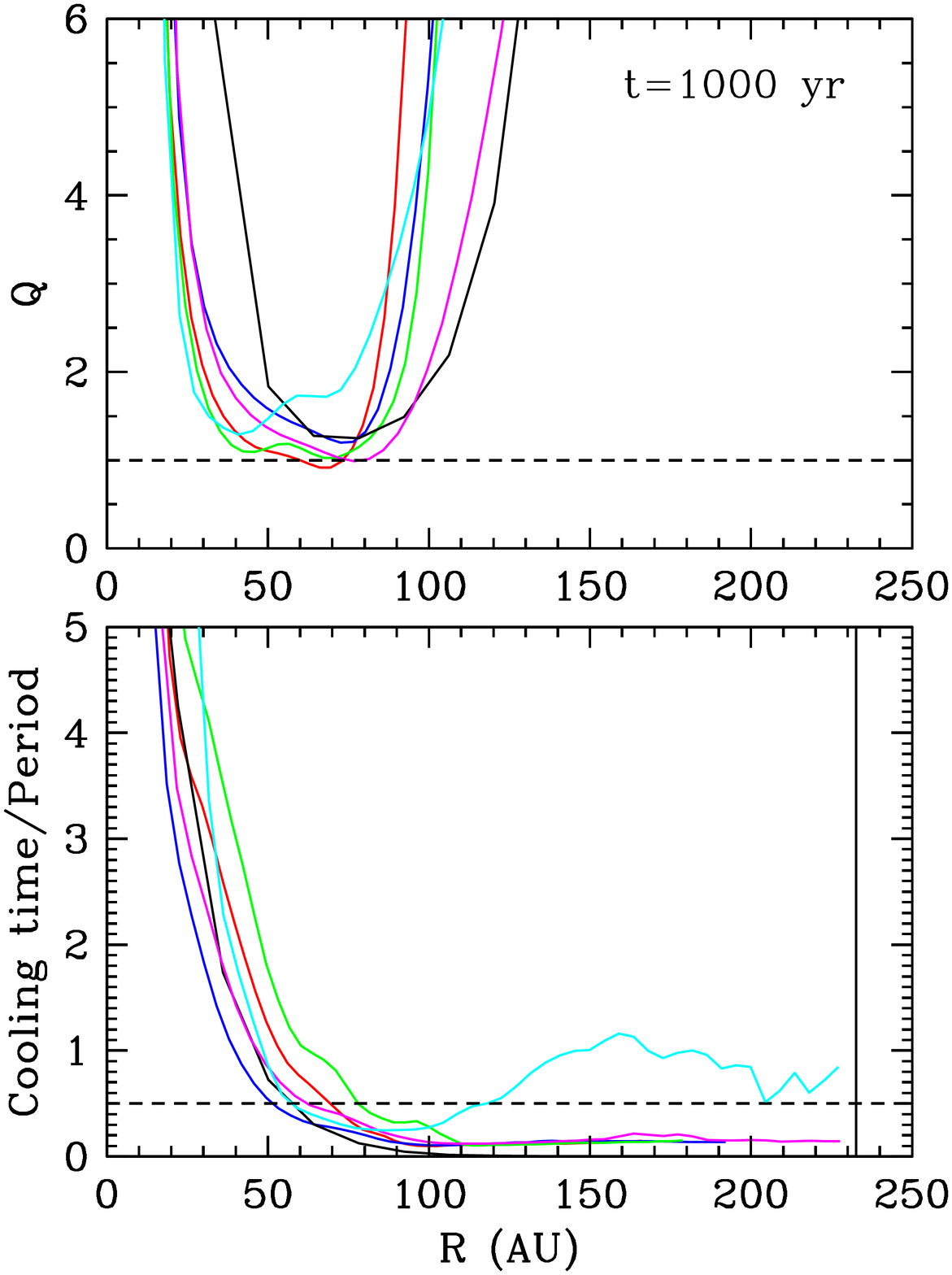}}
\caption{Toomre parameter (top) and cooling time in units of the local orbital period (bottom) for discs that do not fragment, at 1000~yr from the beginning of the simulation.  Both are azimuthally averaged.
The different colours correspond to different runs as listed in Table~\ref{tab:runs}.}
\label{fig:nonfragment}
\end{figure*}

\begin{figure*}
\centerline{
\includegraphics[width=0.9\columnwidth]{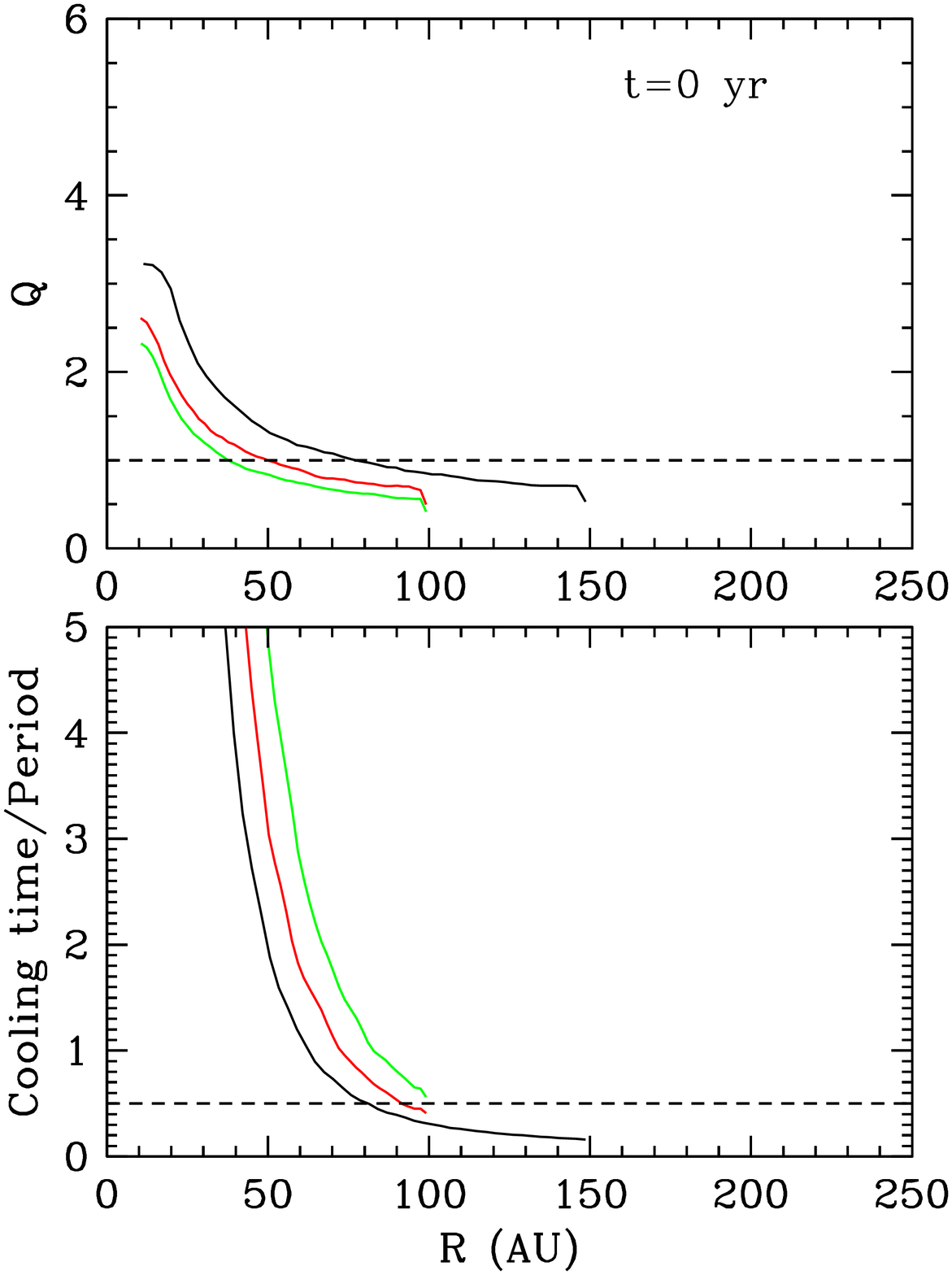}
\includegraphics[width=0.9\columnwidth]{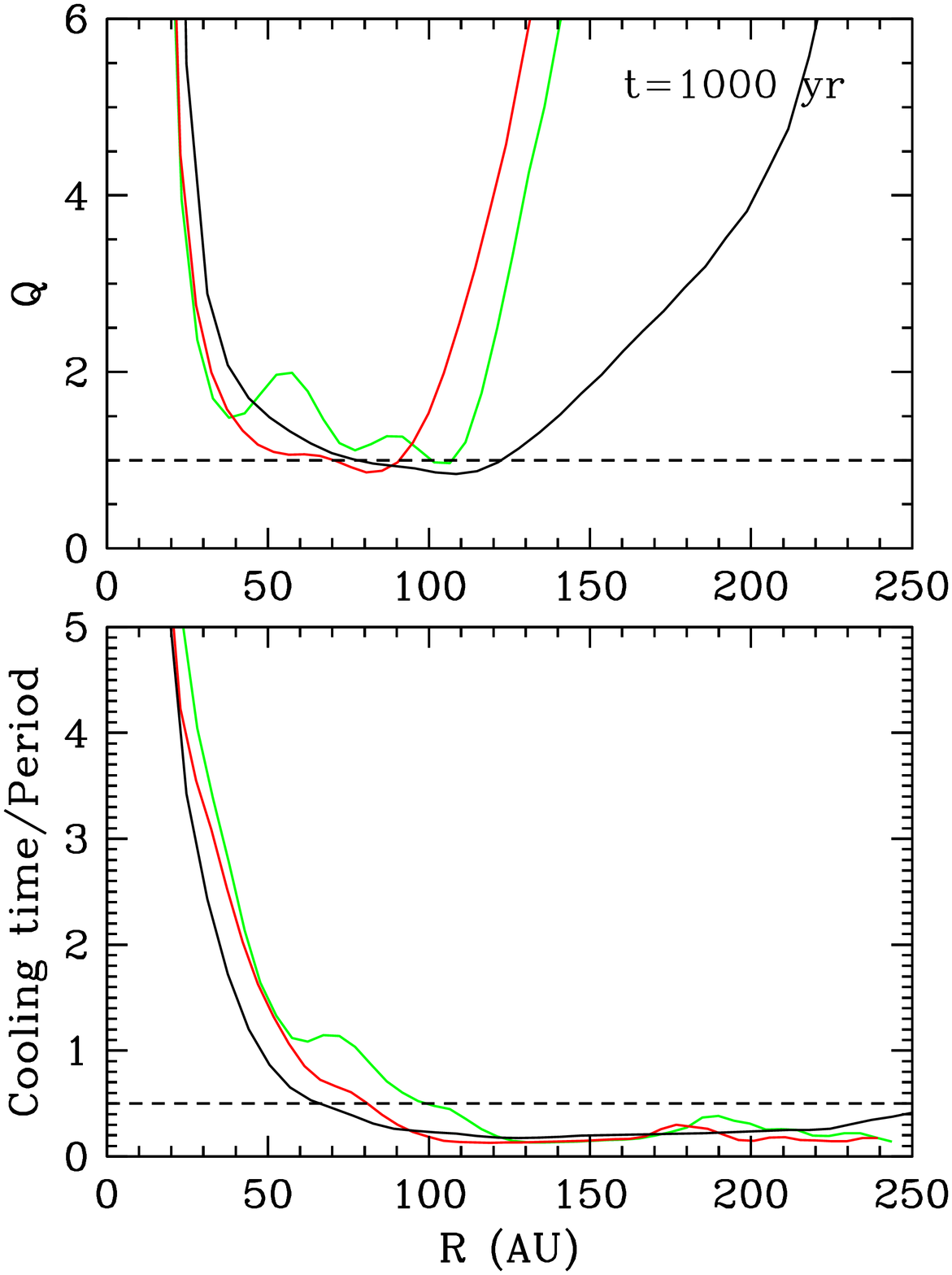}}
\caption{Toomre parameter (top) and cooling time in units of the local orbital period (bottom) for discs that fragment, at 1000~yr from the beginning of the simulation. Both are azimuthally averaged. The different colours correspond to different runs as listed in Table~\ref{tab:runs}.}
\label{fig:fragment}
\end{figure*}

The gas is assumed to be a mixture of hydrogen and helium. We use an EOS (Black \& Bodenheimer 1975) that accounts (i) for the rotational and vibrational degrees of freedom of molecular hydrogen, and (ii) for the different chemical states of hydrogen and helium. We assume that ortho- and para-hydrogen are in equilibrium. For the dust and gas opacity we use the parameterization  by Bell \& Lin (1994), $\kappa(\rho,T)=\kappa_0\ \rho^a\ T^b\,$, where $\kappa_0$, $a$, $b$ are constants that depend on the species and the physical processes contributing to the opacity at each $\rho$ and $T$. The opacity changes due to ice mantle melting, the sublimation of dust, molecular and H$^-$ contributions,  are all taken into account. 

Sinks are invoked in the simulations in order to avoid very small timesteps when the density becomes high. Sinks are created when a particle reaches a critical density of   $\rho_{\rm sink}=10^{-9}$g cm$^{-9}$ and the particles in its neighbourhood (i.e. its 50 neighbours) are bound. The sink radius is set to 1~AU. Once a particle is within this radius and bound to the sink then it is accreted onto it.

The disc is represented by $1.5\times10^{5}$  SPH particles, which means that the minimum resolvable mass (corresponding to the number of neighbours used, i.e. 50 SPH particles) is $\approx 0.1\,{\rm M}_{_{\rm J}}$ for the most massive disc simulated ($M_{\rm D}=0.3~{\rm M_{\sun}}$).  Meru \& Bate (2010) find that disc fragmentation simulations  have probably not converged using $2\times10^{6}$ particles. However, their study is based on a parameterized cooling prescription, and on only two high resolution (with $1.6\times10^{7}$ particles) simulations; more work is needed to establish the effect of resolution in disc fragmentation. Stamatellos \& Whitworth (2009b) have shown that $1.5\times10^{5}$ SPH particles are adequate to properly resolve the Jeans mass, the Toomre mass, and the vertical structure of the disc (Nelson 2006). Additionally, Stamatellos \& Whitworth (2008) demonstrate that, with $1.5\times10^{5}$ particles, the prescription for the radiative transfer used in this work, reproduces well the vertical disc temperature profile derived analytically by Hubeny (1990).

\subsection{The lower limits of disc fragmentation}

We examine the possibility of fragmentation for discs with masses from 0.15 to 0.30 ${\rm M}_{\sun}$ and radii from 60 to 150~AU (around a 0.7~${\rm M}_{\sun}$ star) (see Table~\ref{tab:runs}). All the discs are Toomre unstable or close to being so (Figs.~\ref{fig:nonfragment}, \ref{fig:fragment}). Gravitational instabilities develop and grow in all simulations;  in some of the simulations these instabilities become stronger leading to disc fragmentation within a few thousand years (Fig.~\ref{fig:faceon1}, first column), and in other simulations they weaken resulting in an almost axisymmetric disc in a self-regulating state (Lodato \& Rice 2004, 2005). The simulations are followed for long enough time  (40~kyr or equivalently $\sim 20$ orbital periods at 150 AU) to ensure that gravitational instabilities have adequate time to grow. We find that even a disc with mass of 0.25~M$_{\sun}$ and radius  of $100$~AU fragments; these values and the ratio of disc-to-star mass ($M_{\rm D}/M_\star\approx0.36$) are considerably lower than the ones assumed by Stamatellos \& Whitworth (2009a) (0.7~M$_{\sun}$, 400~AU, and ${M_{\rm D}/M_\star=1}$ respectively). 

 A common characteristic of all simulations is that the the gravitational instabilities provide an effective way to redistribute angular momentum. As a result the discs spread from an initial size of 60-100 AU to more than 200 AU. The discs can cool efficiently only outside 70-100 AU (e.g. Stamatellos \& Whitworth 2009b). If the disc remains Toomre unstable as it spreads then it fragments (runs 1, 3, 5; Fig.~\ref{fig:fragment}). For example, in run 5 (see Table~\ref{tab:runs}), the disc has an initial size of 100 AU, but within 1000 yr its size has reached 250 AU. Subsequently, the disc fragments to form 4 bound objects at distances 100, 140, 175, and 200 AU (see Table~\ref{tab:statistics}). A similar process of disc expansion and fragmentation is suggested by Clarke (2009)  but in her work the disc spreading  happens on a viscous timescale, i.e. it takes much longer (a few $10^5$ yr). 

Typically the outcome of  disc fragmentation are multiple low-mass objects, i.e. low-mass stars, BDs and planetary-mass objects (see Table~\ref{tab:statistics}). Similar to the Stamatellos \& Whitworth (2009a) simulations, the object that forms first migrates inwards in the inner disc region and increases in mass by accreting material, to eventually become a low-mass hydrogen-burning star. Objects that form later,  farther out in the disc, have lower mass, i.e they are usually BDs and planetary-mass objects. In the three cases of fragmenting discs 1, 3 and 4 objects form in each disc, respectively. Where multiple objects form in the disc, they dynamically interact with each other and the central star, and some of them are ejected from the system. This is what happens to the only planetary-mass object formed in the simulations. These results are qualitatively similar to the ones of Stamatellos \& Whitworth (2009a), but a larger ensemble of simulations is needed for a more quantitative comparison. The main difference between the two sets of simulations is that fewer objects are formed in smaller discs (i.e. $1-4$), as compared with larger discs (i.e. $5-12$, Stamatellos \& Whitworth 2009a).

\begin{figure}
\centerline{
\includegraphics[width=8cm]{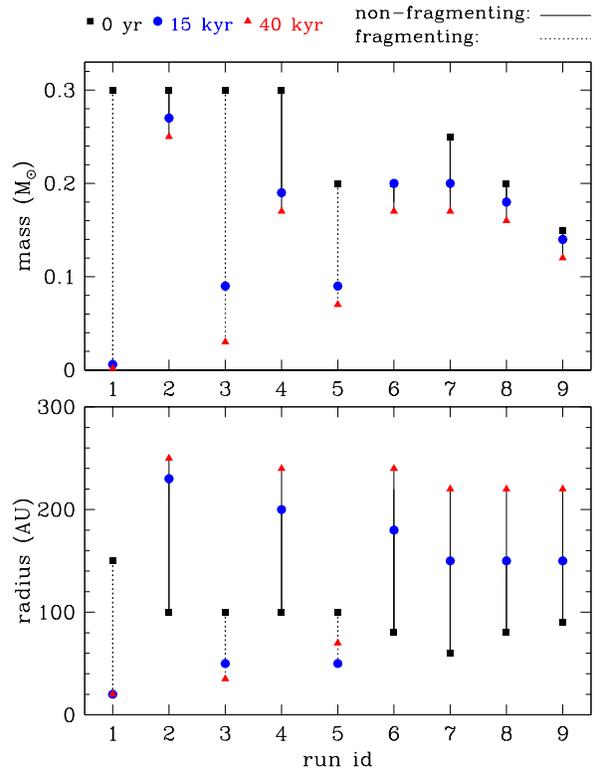}}
\caption{Disc mass (top) and disc radius (bottom) for fragmenting (dashed lines) and non-fragmenting discs (solid lines) at 3 times (0~kyr: black/square, 15~kyr: blue/circle, 40~kyr: red/triangle). The fragmenting discs reduce in mass and size quickly, whereas the non-fragmenting discs expand and dissipate slowly (see text).}
\label{fig:discproperties}
\end{figure}

Following fragmentation, the disc around the central star is considerably reduced in mass and size (Table~\ref{tab:runs}, Fig.~\ref{fig:discproperties}), because a fraction of its mass is going into the new objects that form in its outer regions. The three fragmenting discs  (runs 1, 3, 5) end up with  mass  $0.002-0.07$~M$_{\sun}$, and  size $20-70$~AU, after only 40 kyr (Table~\ref{tab:runs}, Fig.~\ref{fig:discproperties}).  On the other hand, in  discs that do not fragment  the redistribution of angular momentum results in the inner disc regions accreting onto the star and the outer regions expanding (Lynden-Bell \& Pringle 1974). Initially this happens fast since the gravitational torques provide a sufficient effective viscosity, but then it slows down as the disc spreads and becomes more stable. This process is probably not affected by the presence of magnetic fields as the magneto-rotational instability that can provide an additional source of viscosity is presumed to be activated only in the inner disc regions  (within a few AU from the central star; e.g. Zhu et al. 2009). 


\begin{figure*}
\centerline{
\includegraphics[width=0.75\textwidth]{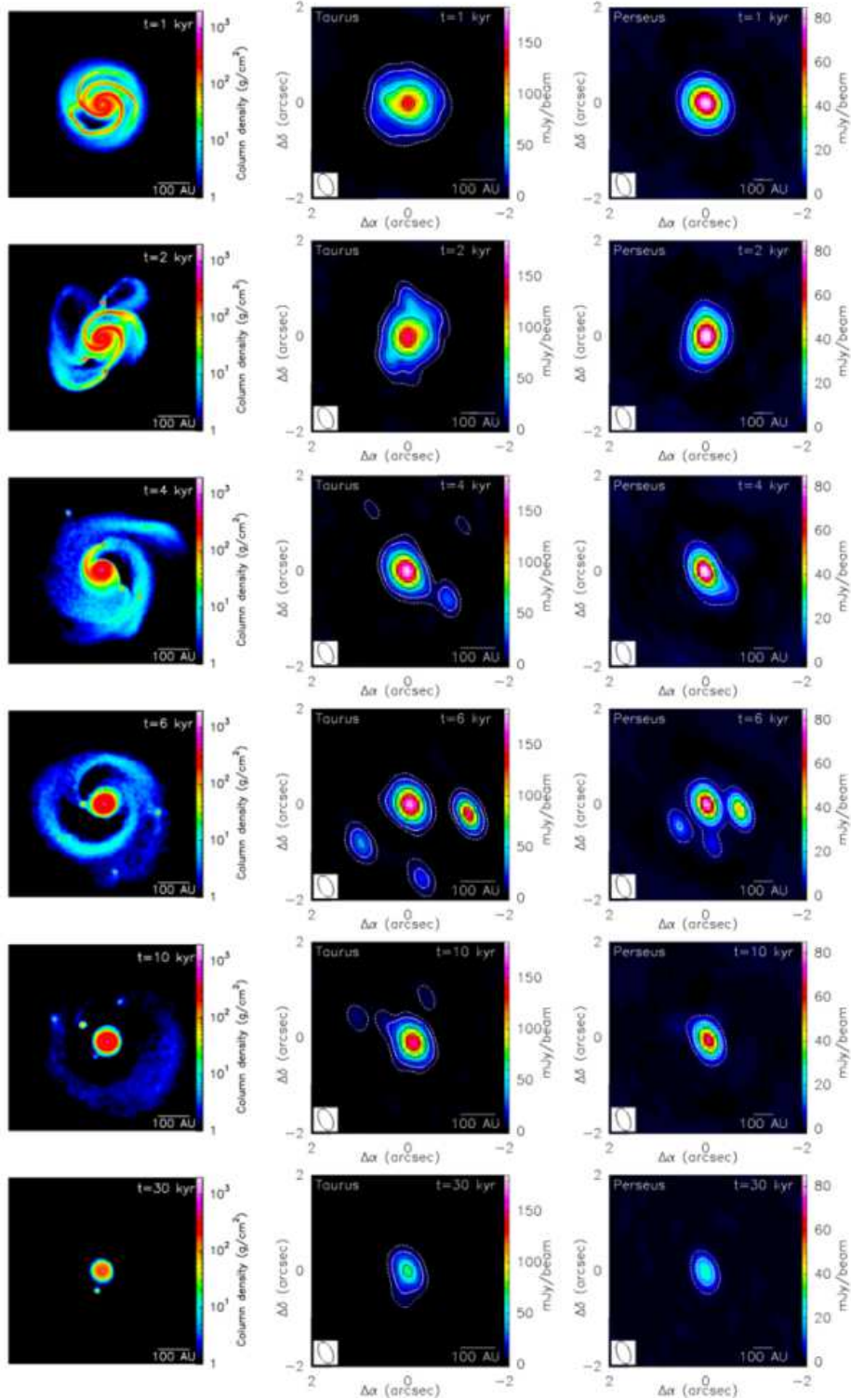}}
\caption{Hydrodynamic simulations (first column) and simulated 1.3-mm observations of fragmenting discs at the distance of Taurus (140~pc; second column), and Perseus (250~pc; third column). 6 snapshots are plotted. The simulations depict the logarithm of column density (g~cm$^{-2}$). The  synthetic images refer to flux density (mJy/beam).  The dashed contours on the synthetic images show the typical  5$\sigma$ detection level ($\sim$3 mJy/beam) that can be achieved by the PdBI (A configuration, i.e. PdB-A). The white contours are levels of 20$\sigma$, and the following black contours correspond to  50, 100 and 200$\sigma$ levels. The ellipse in the bottom-left corner of each simulated image represents the FWHM beam size (resolution element) obtained with PdB-A.}
\label{fig:faceon1}
\end{figure*}

\subsection{Even lower limits for fragmentation?}
\label{subsec:evenlower}

The simulations presented here explore only a small fraction of the parameter space. The criteria for disc fragmentation could be satisfied for even smaller disc masses and sizes. More specifically, the limits for fragmentation could be lower in the following cases.
\begin{enumerate}
\item For a less steep disc density profile. We have assumed that $p=1$, but observations of  T Tauri discs suggest that $p$ could be as low as $\sim\!0.4$ (see Section~\ref{subsec:ics}). This means that there could be more mass at larger  disc radii.  The value of $p$ for early-stage discs is uncertain.
\item For a steeper disc temperature profile. We have assumed (in most simulations) that $q=0.75$, whereas  observations of  T Tauri discs suggest that $q$ could be as high as $\sim\!0.8$ (see Section~\ref{subsec:ics}).  This results in  cooler, and hence more unstable discs. The value of $q$ for early-stage discs is also uncertain.
\item For a lower mass central protostar. We have assumed $M_\star=0.7~M_{\sun}$, but at the initial stages of their formation  protostars are less massive. This has two consequences: (a) their discs could be unstable even if they have masses  below $\sim\!0.01~M_{\sun}$ (e.g. Inutsuka et al. 2010), and (b) their discs can cool fast enough even if they are small, as the minimum radius for disc fragmentation scales as $M_{\star}^{1/3}$ (Whitworth \& Stamatellos 2006; Clarke 2009).
\item  For discs that are perturbed by passing by stars. It has been shown that such encounters promote disc fragmentation (Thies et al. 2010).
\end{enumerate}

Thus, the limits (mass, size) of disc fragmentation are most probably lower than the ones reported here. In a future publication we will estimate the disc fragmentation limits considering the above factors.

\section{Synthetic images of fragmenting discs}
\label{sec:synthetic}

Comparing the results of  current interferometric millimetre observations of young protostars (Class~0, Class I objects) with the predictions (e.g. mass and size of the discs, multiplicity of the stars formed) of the simulations is important for understanding the first steps of low-mass star formation by disc fragmentation.  In the following paragraph we  describe the method used to produce synthetic images from numerical simulations of fragmenting discs, so as to determine whether such discs and their spiral structure are detectable by current facilities, and to compare  these discs with observations.

 The column density maps produced by the numerical simulations are scaled to the distance  of the Taurus (140~pc) and Perseus (250~pc) star forming regions. Then, they are transformed  to flux density images (in mJy/beam) assuming typical dust properties for Class~0 objects: optically thin emission at 1.3~mm, a dust temperature $T_{\rm dust} = 10 $~K and a dust emissivity $\kappa_{\rm1.3mm} = 0.01\, {\rm cm}^2 {\rm g}^{-1}$. (Note that  the dust temperatures may be higher and that the dust opacity is uncertain by a factor of at least $\sim 2$). Finally, the resulting images are adapted to the GILDAS\footnote{Grenoble Image and Line Data Analysis System, software provided and developed by IRAM ({\url{http://www.iram.fr/IRAMFR/GILDAS}})} data format, in order to be processed with the PdBI simulator, mostly comprising the 'uv$\_$model' and 'uv$\_$map' tasks. These tasks allow us to produce simulated 1.3~mm continuum images, i.e. images that would be obtained if these objects were observed with PdBI.  During this last step, we first create the synthetic reference uv-coverages corresponding to the PdBI in its most extended (A) configuration. Pointing towards the coordinates of the source is also considered; then  'uv$\_$model' convolves the flux density maps with the reference uv-coverages, and finally 'uv$\_$map' inverts the resulting uv-tables to create the synthetic 1.3~mm continuum maps.

 Fig.~\ref{fig:faceon1} shows the synthetic images for 6 snapshots of simulation 3 (see Table~\ref{tab:runs}) (second and third column) together with the logarithmic column density maps from the simulation (first column).  The images are 1.3~mm continuum images corresponding to  simulated observations using the A-array of the PdB interferometer at  Taurus (second column) and Perseus (third column). The dashed white contours in the image correspond to the  typical $5\sigma$  level   ($\sim\!3$~ mJy/beam) that can be achieved in the 230 GHz continuum with the A-array of PdBI, taking into account the increased sensitivity of the newly installed WideX correlator.  We see  that the PdBI is sensitive enough to detect  fragmenting discs as extended structures  (expected angular diameter $\sim$ 2$\arcsec$-3$\arcsec$ compared to a PdBI beam size $\sim$ 0.5$\arcsec$ in the A configuration). On the other hand, PdBI  does not provide enough resolution to discern any spiral structure in the disc. However,  the upcoming ALMA should in principle be able to detect such spiral structures  out to distances of the Taurus-Auriga star-forming regions, i.e. at $\sim$~140~pc (Cossins et al. 2010).  
 
 The disc around the protostar  (Fig.~\ref{fig:faceon1}) initially is large (100~AU)  and massive ($0.3~{\rm M}_{\sun}$) but within a few thousand years it reduces in size ($\sim35$~AU) and mass ($\sim0.03~{\rm M}_{\sun}$), as it fragments to form companions (i.e. low-mass stars and BDs). Thus, though fragmenting discs can in principle be detected, they may be rarely observed  due to their short lifespan (see next Section).  Companions to the central protostar (which are signposts of fragmentation) can be detected if they are attended by discs that have enough mass. When such discs are disrupted (e.g. by dynamical interactions with other companions in the disc) then the low-mass companions may remain undetected (e.g. 4$^{\rm th}$ row, 3$^{\rm rd}$ column). They also have to be away from the central protostar, so that they can be resolved as separate entities (e.g. in 5$^{\rm th}$ row, 2$^{\rm nd}$ column, the close companion is blended with the central protostar).

\section{Comparison with disc observations}
\label{sec:compare}

 \subsection{Uncertainties in calculating disc masses}
 \label{subsec:uncertainties}
The masses and radii of early stage discs are difficult to determine. They are rather uncertain, maybe up to an order of magnitude (see discussion in J{\o}rgensen et al. 2009). Discs around newly formed protostars (i.e. Class 0) are generally hidden within the envelope that surrounds the protostar, so that radiative transfer models are needed in order to distinguish the submm and mm emission of the disc from the emission of the envelope. For example, the J{\o}rgensen et al. (2009) models suggest that  in Class 0 objects  68\% of the observed interferometric flux is from the disc (87\%  in Class I objects). Once the  integrated  flux, $F_{\rm int}$, from the disc is known then its mass is computed using 
\begin{eqnarray}
M_{\rm D} &=& 4.0\times10^3\  {\rm M}_{\sun} 
\left(\frac{F_{\rm int}}{10~{\rm mJy}}\right) \nonumber
 \left(\frac{d}{{\rm 140~pc}}\right)^2 \times \\
& &\times  \left(\frac{\kappa_{\rm 1.3mm}}{0.01\ {\rm cm}^2~{\rm g}^{-1}}\right)^{-1}
\left(\frac{<T_{\rm dust}>}{20~{\rm K}}\right)^{-1}\,.
\end{eqnarray}
Typical values for the dust temperature and the opacity are $<T_{\rm dust}>=20~$K and $\kappa_{\rm 1.3mm}=0.01\ {\rm cm}^2~{\rm g}^{-1}$, respectively but they are rather uncertain. Ossenkopf \& Henning (1994) estimate that opacities may increase as much as an order of magnitude depending on the density. The dust temperatures in discs are also poorly constrained and the use of an average dust temperature may be inappropriate.

\subsection{Candidate unstable discs}

The fragmenting discs (runs 1, 3, 5; Table~\ref{tab:runs})  in our simulations reduce  to a mass of $0.002-0.07~{\rm M}_{\sun}$, and a size of $20-70$~AU, within 40~kyr (see  Fig.~\ref{fig:discproperties} and Table~\ref{tab:runs}). This happens so fast, that large, unstable discs or discs that are currently undergoing fragmentation should be rarely observed (see Discussion). 
There are only  a few observations of discs that (considering the uncertainties in the estimated disc masses and radii, see Section~\ref{subsec:uncertainties}, and in disc fragmentation models, see Section~\ref{subsec:evenlower}), may be gravitationally unstable. A recent example is the disc around HL Tau, a Class II object, which has a mass of around 0.1 M$_{\sun}$ and extends to at least 100 AU, with a possible low-mass fragment at $\sim 65$~AU (Greaves et a. 2008). It has been shown that this fragment may have formed by disc fragmentation (Greaves et a. 2008). Another example of a possible unstable disc, is Serpens FIRS 1 with mass of 1.0~M$_{\sun}$ and radius 300~AU (Enoch et al. 2009a); this however may be a rotating collapsing spheroid. Other discs that may be gravitationally unstable are GSS39 (Class II; 0.143~M$_{\sun}$, 200~AU) and DoAr 25 (Class II: 0.136~M$_{\sun}$, 80~AU) in Ophiuchus  (Andrews et al. 2009), and  HC189 (0.25~M$_{\sun}$, $<100$~AU) in the Orion Nebula Cluster (Eisner et al. 2008). Another candidate is the disc  around IRAS 16293-2422B (Class 0; Rodriguez et al. 2005) which is resolved with the   VLA  and has an estimated mass of  $0.3-0.4\ {\rm M}_{\sun}$ and size of 26~AU. 

\subsection{Remnants of discs that may have fragmented}

Discs that have masses and radii consistent with being remnants of discs that have fragmented in the past, are routinely observed at the Class I and Class II stages; however,  it is difficult to establish whether fragmentation has occurred. Discs with masses and sizes consistent with the final ($t=40$~kyr) masses and sizes in our simulations ($0.002-0.07~{\rm M}_{\sun}$, $20-70$~AU) have been observed  (e.g. Eisner et al. 2005, 2006; 2008; Rodriguez et al. 2005; Eisner \& Carpenter 2006; Brinch et al. 2007; Andrews \& Williams 2007; Andrews et al. 2009; J{\o}rgensen et al. 2009; Brinch et al. 2009; Tobin et al. 2010). 

\subsection{Surveys of early-stage discs}

Observations of early-stage discs are difficult as these discs are deeply embedded in their parental clouds. In the following paragraphs we discuss a few surveys that report relatively massive disc candidates.

J{\o}rgensen et al. (2009) using submm/mm observations and radiative transfer modelling computed  disc masses  from 0.017 to 0.46~M$_{\sun}$, and from 0.008 to 0.053~M$_{\sun}$ in a sample of Class 0 and Class I sources, respectively. All of these discs were unresolved or marginally resolved (size $\stackrel{<}{_\sim} 1\arcsec-3\arcsec$, i.e. $\sim 125-700$~AU).  In a few of the Class I sources of the J{\o}rgensen et al. sample HCO$^+\ 3-2$ line observations revealed signs of Keplerian rotation confirming their disc status (e.g. L1489-IRS; Brinch et al. 2007).  Scattered light images of L1527 and radiative transfer modelling infer a disc with radius $\sim 190$~AU and mass of 0.005~M$_{\sun}$  (Tobin et al. 2010). We note however that the mass of this disc could be up to 0.1 ~M$_{\sun}$ if different opacities are assumed (Loinard et al. 2002). Interestingly, a close companion at projected distance of $\sim 24$~AU is also present (Loinard et al. 2002),  indicative of disc fragmentation. In the same sample of J{\o}rgensen et al. (2009), NGC1333-IRAS2 (Class 0) is estimated to have a radius  of $R=150$~ AU and mass of  $M=0.1$~M$_{\sun}$ although no sighs of Keplerian rotation were found (J{\o}rgensen et al. 2005; Brinch et al. 2009).
The estimated masses and sizes L1527 and NGC1333-IRAS2 are contested as  recent 1.3-mm observations (Maury et al. 2010) reveal no discs larger than the resolution limits of PdB-A ($\sim100$~AU).

Eisner et al. (2005, 2006, 2008) observed young protostars (Class 0s and Class Is) in the Orion Nebula Cluster and using radiative transfer models inferred  discs with masses up to 0.39~M$_{\sun}$ and radii up to 220~AU.  

Andrews et al. (2009) used the SMA to achieve high angular resolution of $\sim0.3\arcsec$, i.e.  $\sim40$~AU in Ophiuchus  and observed discs with masses up to 0.14 M$\sun$ sizes up to 200 AU; some of them show signs of Keplerian rotation (evident in CO~${j=3-2}$ line observations).  In the Taurus-Auriga  and Ophiuchus-Scorpius star forming regions Andrews \& Williams (2007) found disc masses of up to 0.17~M$_{\sun}$ and sizes up to 700~AU (most of these around Class II objects).
 
\subsection{Remarks}
Discs are probably present at the earliest phases of star formation, i.e. in Class 0 and Class I objects. However, their masses and sizes are debated. There are a few candidates that they may be relatively massive (a few 0.1~M$_{\sun}$) and large ($\sim100$~AU), but current observations are not conclusive (e.g. see Maury et al. 2010). Comparing typical  masses and radii  derived from disc observations with those of fragmenting discs in Table~\ref{tab:runs} (see also Fig.~\ref{fig:discproperties}) we find that most of these discs are not gravitationally unstable. However, it cannot be excluded that these discs were gravitationally unstable and that have fragmented in the past (i.e. in last $\sim40$~kyr).






\section{Discussion}

The ensemble of disc simulations presented here show that protostellar discs fragment when they are large enough ($\stackrel{>}{_\sim} 100$~AU) so that their outer regions can cool efficiently, and massive enough to be gravitationally unstable, at such radii.  We find that even a disc with moderate size (100~AU) and moderate mass ($0.25~{\rm M}_{\sun}$) around a $0.7$-${\rm M}_{\sun}$ protostar ($M_{\rm D}/M_\star\approx0.36$) fragments. These values are considerably smaller than those in previous studies; Stamatellos \& Whitworth (2009a) simulated discs with size 400~AU and mass $0.7~{\rm M}_{\sun}$, around a $0.7$-${\rm M}_{\sun}$ protostar ($M_{\rm D}/M_\star=1$). Even less massive and smaller discs may be able to fragment (see Section~\ref{subsec:evenlower}); a larger parameter space needs to be explored in order to determine the lowest limits of disc fragmentation.

The properties of the objects formed by the fragmentation of the  discs presented in this paper show similarities with the properties of the objects produced in  larger, more massive discs (Stamatellos \& Whitworth  2009a).  Most of the objects formed are BDs, but low-mass H-burning stars, and planetary-mass objects also form.  Low-mass H-burning stars generally orbit closer to the central star, BDs are on wider orbits ($170-1250$~AU), whereas planetary-mass objects tend to be ejected from the disc, due to dynamical interactions. The most notable difference is that fewer objects form in smaller discs. A larger ensemble of simulations is needed to make a more quantitative comparison with the results of Stamatellos \& Whitworth~(2009a).

Fragmenting discs can in principle be detected by current (e.g. PdBI; Section~\ref{sec:synthetic}) and future (ALMA; Cossins et al. 2010) interferometers. However, the chance to observe such discs during fragmentation is small, as fragmentation happens in a dynamical timescale, i.e. within a few thousand years.   Hence, a large number of protostars need to be observed. Based on first results (Andr\'e et al. 2010), the total number of Class~0 protostars  that are expected to be found in nearby clouds by the Herschel Gould Belt survey is $\stackrel{>}{_\sim}350$. Assuming that (i) all protostars are attended by discs that grow, become unstable, and fragment during the Class~0 phase, (ii) the lifetime of the Class 0 phase is $\sim1.5\times10^5$ yr (Evans et al. 2009; Enoch et al. 2009b), and (iii) in the disc fragmentation scenario the disc fragments and therefore dissipates within $1.5\times10^4$ yr (see Fig.~\ref{fig:discproperties}, red triangles), it follows that $\sim\!10\%$ of Class 0 objects could have extended unstable discs.  However, Stamatellos \& Whitworth (2009a) find that typically a few BDs form in each fragmenting disc around a  Sun-like star, and suggest that only  $\sim20\%$ of Sun-like stars need to have  unstable extended discs to produce all BDs and a large fraction of low-mass hydrogen-burning stars. Hence, assuming that disc fragmentation delivers half of the observed BDs, i.e. that  only $\sim10\%$ of Sun-like stars have discs that fragment, then the percentage of  Class 0 objects with unstable extended discs is only  $\sim\!1\%$. This percentage could be even lower if a smaller fraction of BDs forms by fragmentation of early stage discs or if discs need not be that extended to fragment (see Section~\ref{subsec:evenlower}). On the other hand it could be higher if the lifetime of the Class 0 phase is shorter and the disc dissipation timescale longer. For example, assuming a Class~0 lifetime of $\sim3\times10^4$ yr (Andr\'e et al. 2000) and a similar disc dissipation timescale ($\sim 3\times10^4$ yr) then  the percentage of Class~0 objects with large extended discs is $\sim\!10\%$ (for half of the observed BDs to form by disc fragmentation). These estimates may also depend on how fast a disc assembles around a young protostar forming in a collapsing molecular cloud core; larger scale simulations (Walch et al. in prep.), including the effects of radiative transfer and magnetic fields (e.g. Commer{\c c}on  et al. 2010), are needed for assessing this.

It is possible to observe discs that are remnants of disc that have fragmented in the past. Our simulations show that such discs may reduce to sizes  $\sim20-70$~AU and masses  $\sim 0.002-0.07~{\rm M}_{\sun}$, within only 40~kyr. A comparison with observations shows that there are  discs with similar properties (see Section~\ref{sec:compare}). However, it is not possible to establish with certainty that fragmentation has occurred. The presence of companions may be indicative of fragmentation, but these companions (i) may be too close to the central protostar to be resolved as separate objects, (ii) may be too far away from the central protostar (e.g. within a few thousand AU) to be observed and/or associated with it, and (iii) may be below the sensitivity limit if they are not attended by an envelope/disc with enough mass to emit considerably in submm/mm wavelengths. 

Discs that do not fragment dissipate on a viscous timescale, i.e. on the order of 1~Myr.   For example the non-fragmenting discs  in our simulations (Fig.~3 and Table~\ref{tab:runs}) lose only a small fraction of their mass (as material is accreted onto the protostar) and increase in size as angular momentum is redistributed within the disc. Hence, if  massive ($\stackrel{>}{_\sim}0.2$  M$_{\sun}$), extended ($\stackrel{>}{_\sim}100$~AU)  discs do not fragment then they should be readily and routinely observed, which is not the case. Therefore such discs either form and quickly fragment after becoming unstable due to mass loading from the surrounding envelope, or that their formation is suppressed (e.g. by magnetic fields).

We conclude that current disc observations cannot exclude the mechanism of disc fragmentation as a means of producing low-mass hydrogen burning stars, brown dwarfs, and planetary mass objects. Interferometric observations of larger samples of Class 0 objects with, e.g.,  ALMA may however
provide a much stronger test of this scenario in the future.

\section*{Acknowledgements}
Simulations were performed using the Cardiff HPC Cluster {\sc Merlin}. Colour plots were produced using {\sc splash} (Price 2007). DS and APW acknowledge support by the STFC grant ST/HH001530/1. 
The work presented in this paper was stimulated by discussions held in the context of the Marie Curie Research Training Network ÒConstellationÓ (MRTN-CT2006-035890).

\label{lastpage}
\end{document}